\documentstyle[12pt,twoside,fleqn,espcrc1]{article}
\title{
Microscopic description of light unstable nuclei with the 
stochastic variational method}
\author{
   K. Varga\address{
   Institute of Nuclear Research of the Hungarian \\
   Academy of Sciences, Debrecen, H-4001, Hungary},
   Y. Suzuki\address{
                  Department of Physics, Niigata University \\
   Niigata 950-21, Japan},
   K. Arai\address{Graduate School of Scinece and Technology,\\ 
   Niigata University, Niigata 950-21, Japan}, and
   Y. Ogawa\address{RIKEN, Hirosawa, Wako, Saitama 351-01 Japan  }}
\begin{document}
% typeset front matter
\maketitle

\begin{abstract}
\noindent
The structure of the light proton and neutron rich nuclei is studied
in a microscopic multicluster model using the stochastic variational 
method. This approach enables us to describe the weakly bound nature 
of these nuclei in a consistent way. Applications for various nuclei
$^{6-9}$Li, $^7$Be, $^8$B, $^9$C, $^{9-10}$Be, $^{9-10}$B presented.
The paper discusses the relation of this model to other models as well as
the possible extension for p and sd shell nuclei.

\end{abstract}

\section{Introduction}
The experimentally intensively studied unstabil nuclei \cite{exp}
have challenged the theoretical nuclear physicists. These nuclei
exhibit a ``nucleon halo'', a new form of nuclear matter,
characterized by low and nonuniform density distribution. To describe 
this nucleon (proton or neutron) halo care must be taken on the proper
description of the nuclear dynamics. The halo stucture is mostly 
interpreted as two- or three-body phenomena, that is most of the 
theoretical models assume a stabil core and one or two valance nucleons
\cite{NBR}. Although one can understand various properties of the
halo nuclei in these phenomenological models, it is a natural aim 
to go beyond these approaches. While there are sophisticated techniques
to tackle the two- and three-body problems, the solution of the nuclear
many-body problem is still much too complicated. The small separation energies
a characteristic property of the halo nuclei, require extra caution. 
One often has to reproduce as small as 100 keV energy differences (for example 
the proton separation energy of $^8$B or neutron separation energy of 
$^{11}$Li). These tiny energies play a crucial role in determining the asymptotic
form of the wave function and therefore the halo structure and the physical
properties characterizing the halo nuclei (momentum distribution, interaction 
cross section) depends on them very much. A reliable description, therefore,
should be able to give accurate energies and correct asymptotic behaviour. 
Another important feature to be taken into account is the presence of correlations 
between nucleons. While the correlation of nucleons is a well known property in
nuclear physics, it is somewhat surprising that there are reasonable experimental
indications to assume the existence of correlated neutron pairs in such a low
density nucleon matter as the halo of $^6$He or $^{11}$Li. The above metioned 
specific conditions (correlation, extended nonuniform density distribution, 
importance of asymptotic part) make very difficult the application of models using 
single particle bases. Much of the succes of the three-body desription is 
due to their capability to treat these requirements  by applying a wave function 
depending on {\sl relative} variables. The antisymmetrization is, however, 
much more difficult in the case of relative coordinates and  one often sacrifices 
the exact treatment of the Pauli principle in favor of simpler model, assuming 
structureless clusters or ``core'' (for example $^9$Li+n+n). At this point,
besides the validity of assuming a  simplified core, another problem, 
the question of interaction between constituent particles appear. 
\par\indent 
Our approach tries to combine the advantages of different descriptions. A 
microscopic framework is used, that is we treat nucleonic degrees of freedom 
thereby avoiding the need of knowledge of core-nucleon or nucleus-nucleus 
interaction. We use relative coordinates to have a flexible coordinate-system 
to describe the dynamics and to eliminate any problem in connection 
with the center-of-mass motion. Correlated basis functions is used  to 
answer the challanges of the weakly bound halo structure. The spatial part 
of the basis functions are chosen to be Gaussian form to facilitate the 
fully analitical calculation of the marix elements. The Pauli principle is 
treated exactly.

\section{Formalism} 
In our variational approach the basis functions are assumed to have the form 
\cite{vk}
\begin{equation}
\psi_{(LS)JMTM_T}({\bf x},A)={\cal A}\lbrace \phi_{int}{\rm e}^{-{1\over 2} 
{\bf x} A {\bf x}}\left[\theta_L({\bf x})\chi_S\right]_{LM} \eta_{TM_T}
\rbrace ,
\end{equation}
where ${\bf x}=({\bf x}_1,...,{\bf x}_{N-1})$ is a set of relative (Jacobi) 
coordinates, the operator ${\cal A}$ is an antisymmetrizer, the function 
$\theta_{LM_L}(\bf x)$ represents the angular part of the wave function,
$\chi_{SM_S}$  is the spin function and 
$\phi_{int}$ is the intrinsic function of the clusters. The angular part
$\theta_{LM_L}(\bf x)$ is a vector coupled product of spherical harmonics
of the relative coordinates. The spin of the clusters coupled to total spin
S. 
\par\indent
The assumption of clusters is based on physical motivations as well as practical
considerations. The clustering in light nuclei  has long been known and 
supported by various experimental facts as well as numerical simulations 
\cite{AMD,FMD}. If one assumes
Gaussian packet $\varphi^{\gamma}_{\bf s} ({\bf r})=(2\gamma/\pi)^{3/4} 
{\rm exp} \lbrace -\gamma({\bf r}-{\bf s})^2 \rbrace$  
single particle states to describe a nucleus and try to find the optimal positions of the centers
of the Gaussian packet, that is try to find the configurations that minimize the total energy in a realistic potential, one observes the formation 
of various clusters of nucleons (alpha, triton, $^3$He, etc.). The experimental 
and theoretical research  devoted to  unstable nuclei is also in favor of this 
approach (e.g. $^6$He=$\alpha$+n+n, $^{11}$Li=$^9$Li+n+n, and so on). 
The wave function of the clusters are approximated by  a single harmonic
oscillator shell model configuration. If more accurate description of the
internal structure of clusters is needed one can superpose shell model 
configurations \cite{Descouvemont}. Alternatively, one can divide the clusters 
into smaller entities, for example if the description of the triton with a 
single shell model configuration insufficient one can descibe it as $p+n+n$.
The model assumes an equal harmonic oscillator size parameter $\beta$ for the clusters. 
\par\indent
The assumption of clusters with relatively simple description of their 
structure requires the usage of an effective nucleon-nucleon interaction,
because a realistic interaction with a strong repulsive core would not render
the clusters to be bound. The effective interaction used in this calculation
(Minnesota interaction) contains a  spin-isospin dependent central and
a spin-orbit potential. The Coulomb interaction between protons is treated 
exactly. 
\par\indent
The matrix elements appearing in the variational equation are calculated
in a fully analytical way \cite{vk}. 

\section{The stochastic variational method}

The adequate choice of the nonlinear parameters (the elements of the 
$(N-1)\times(N-1)$ matrix $A$) is very important. 
As the trial function contains a large number of nonlinear parameters 
and one has to superpose 
many basis functions to get the energy minimum a direct optimization 
of the parameters is not suitable. The number of the different spin-isospin 
and partial wave channels further complicates the choice of the basis 
functions. Moreover, the basis functions are nonorthogonal 
and none of them are indispensible, any of them can be equally represented
by some other choice. This property qualifies a random selection of the
basis function by judging on their contribution to the energy. 
We set up the basis stepwise by randomly choosing $A$ from a 
preset domain of the parameter space and increase the basis dimension by one
if the energy gain by including the randomly selected basis element is 
larger than a preset value, $\epsilon$. This  is repeated until the 
energy converges. This random selection of the nonlinear parameters gives
very accurate energy and keeps the size of the basis feasible. We are able to 
solve $N=2-7$ body problems with this simple strategy. The results, as 
shown in Table 1.,  agree very well with those in the literature.
We also mention that our calculations  for different potentials and different
physical systems (including Coulombic systems in atomic physics) proved to 
be as accurate as the most precise methods. The accuracy reached in these 
calculation justifies the application of this method for the field of 
unstable nuclei.
\par\indent
The spin-isospin and partial wave channels are also randomly selected 
eliminating any bias from the construction of the wave function. 
Several examples show that this random selection gives the same percentage
of different components of the wave function as a more direct way 
(e.g. choosing appropriate basis states in all possible channels and 
diagonalizing in the resulted large space) and it does not lead to a false 
wave function.  

\begin{table}[t]
\caption{Energies and rms radii
of $N$-nucleon systems interacting via the Malfliet-Tjon
potential.}
\begin{tabular}{lclllr}
$N$ & $(L,S)J^{\pi}$&Method  & $E$ (MeV) & $\langle r^2\rangle^{1/2}$ (fm) & ${\cal K}$ \\
\hline
\hline
3 &$(0,1/2)1/2^{+}$
      & Faddeev[7]   & $-$8.25273   &        &    \\
  &   & SVM      & $-$8.2527    & 1.682  & 80 \\
\hline
4 &$(0,0)0^{+}$
       &   CRCG[8]  & $-$31.357 &       & 1000 \\
       &  & SVM  & $-$31.360 & 1.4087  &  150 \\
\hline
5 &$(1,1/2)3/2^{-}$  
& VMC[9]     & $-$42.98 & 1.51 &  \\ 
      && SVM                           & $-$43.48 & 1.51 & 500  \\
\hline
6 ($^6$He) & $(0,0)0^{+}$
 & VMC[9] & $-$66.34& 1.50 &  \\ 
           && SVM                         & $-$66.30 & 1.52   & 800   \\
\hline
7 ($^7$Li)& $(1,1/2)3/2^{-}$  
 & SVM                         & $-$83.4 & 1.68    &1300   \\
\end{tabular}
\end{table}

\section{The microscopic multicluster model}

The method has been applied for various nuclei, such as $^{6-8}$He
 \cite{VSL,VSO},
$^{6-9}$Li, $^7$Be, $^8$B, $^9$C, $^{9-10}$Be, $^{9-10}$B \cite{ASV,li9,be9}. 

Recently, we have studied the  mirror nuclei $^9$C and $^9$Li  in a 
microscopic $\alpha+ ^3$He$+p+p$ and $\alpha+ ^3$H$+n+n$ four-cluster 
model \cite{li9}.
The $^7$Be--$^7$Li and $^8$B--$^8$Li mirror two- and three-body subsystems 
are also investigated with the same effective interaction \cite{minnesota}.
The calculated ground state energies, the radii, and the densities of the 
nucleons are in good agreement with the experimental data.
The magnetic and quadrupole moments, except for
the magnetic moments of $^8$B and $^8$Li, are also reproduced well.
The quadrupole moments of $^9$C and $^7$Be
are predicted to be $-5.04$ e fm$^2$ and $-6.11$ e fm$^2$.
The microscopic multicluster model predicts that the neutron skin thickness is
about 0.4 fm in $^8$Li and $^9$Li, while the proton skin thickness is 0.5 fm
in $^8$B and $^9$C. Comparing to the neutron skin thickness of 0.8 fm found in
$^6$He and $^8$He \cite{VSO}, we conclude that these nuclei do
not show pronounced halo structure.

The mirror nuclei $^9$Be and $^9$B have also been described in our model 
\cite{be9}. These nuclei are 
described in a three-cluster model comprising two $\alpha$-particles and 
a single nucleon. The three-body dynamics of the clusters is 
taken into account by 
including both of the possible arrangements, $(\alpha\alpha)N$ and 
$(N\alpha)\alpha$, and by using all the relevant partial 
waves of the relative motion of the clusters.
The ground state of $^9$Be, the only 
particle-bound state in the spectra of these nuclei, is calculated by  
using the stochastic variational method, while the other 
particle-unbound states are studied by the complex scaling method.  
The calculated spectra of $^9$Be and $^9$B are compared with experiment  
in Table 2. The theoretical level sequence in $^9$Be has a good 
correspondence with the observed 
spectrum. The second $\frac{3}{2}^-$ resonance is obtained at 4.3 MeV 
excitation energy. The other calculations 
\cite{okabe,suppl,desc} also predict the $\frac{3}{2}^{-}_{2}$ state. 
Although no such state is cited in Ref. \cite{ajzen}, the calculated 
resonance may correspond to the state at 5.59 
MeV mentioned in Ref. \cite{dixit}.  We get two broad overlapping resonances 
with $\frac{7}{2}^-$ and $\frac{9}{2}^+$ at about 6.5 MeV. This   
agrees with the conclusion of the recent experiments \cite{dixit,glickman}. 
We could not find a resonance with $\frac{1}{2}^-$ around 8 MeV 
excitation energy in accordance with Refs. \cite{dixit,glickman}, 
although such a state is parenthetically quoted 
in Ref. \cite{ajzen}. Instead of this a $\frac{5}{2}^-$ resonance 
is obtained at 7.9 MeV, which agrees with the result of Refs. 
\cite{okabe,suppl}.  The spectrum 
of $^9$B is less known experimentally compared to that of $^9$Be. 
The calculated spectrum is similar to the one of $^9$Be.  
We can predict the energy and the width of several resonances in 
$^9$B with the same accuracy as the case of $^9$Be.  For example, our 
calculation predicts a missing $\frac{1}{2}^-$ state of a 1 MeV width 
at the excitation energy of 2.43 MeV, which is 
in agreement with the result of a recent $^9$Be($p,n$) 
reaction \cite{pugh} that located the $\frac{1}{2}^-$ state at 2.83 MeV. 
Although no definitive spin assignment is made to the state at 2.788 MeV 
excitation energy \cite{ajzen}, our calculation supports a 
$\frac{5}{2}^+$ assignment rather than $\frac{3}{2}^+$.
The first excited 
$1/2^+$ state was not localized in the present study.
\newpage

\begin{table}
\caption{Energies and widths of the unbound states in $^9$Be and $^9$B. 
The energy is from the three-body threshold. The spin and parity of the 
3.065 MeV state of $^9$B is assumed to be $\frac{5}{2}^+$.}
\end{table}

\vspace {2mm}

\begin{tabular}{p{1.5cm}p{1.5cm}p{3.2cm}p{3.2cm}p{2cm}p{2cm}}
\hline\hline

 & & \multicolumn{1}{r}{exp.$^a$}&  &  \multicolumn{1}{r}{cal.} &  \\ \hline

 & $J^{\pi}$ & $E$(MeV$\pm$keV) & ${\it \Gamma}$(MeV$\pm$keV) & 
               $E$(MeV) & ${\it \Gamma}$(MeV) \\ \hline 

 & $3/2^-$ & $-$1.5735  & ------- &  $-$1.431 &  -------  \\

 & $1/2^+$ &  0.111$\pm7$     &  0.217$\pm$10  &           &  \\

 & $5/2^-$  & 0.8559$\pm1.3$   &  0.00077$\pm0.15$ &  0.84  & 0.001  \\

 & $1/2^-$ &  1.21$\pm120$      &  1.080$\pm$110  &  1.20     &  0.46   \\ 

 & $5/2^+$ &  1.476$\pm9$     &  0.282$\pm$11  &  1.98    &  0.6   \\ 

 $^9$Be & $3/2^+$ &  3.131$\pm25$    &  0.743$\pm$55 &  3.3    &  1.6   \\ 

 & $3/2^{-}_{2}$ & 4.02$\pm100^{\:b}$ &  1.33$\pm$360 &   2.9 &  0.8     \\ 

 & $7/2^-$ &  4.81$\pm60^{\:b}$ &  1.21$\pm$230   &  5.03     &  1.2     \\ 

 & $9/2^+$ &  5.19$\pm60^{\:b}$ &  1.33$\pm$90   &  4.9      &  2.9     \\

 & $(1/2^-)$ &  6.37$\pm80$   &  $\sim$1.0  &     &          \\ 

 & $5/2^{-}_{2}$ &      &          &  6.5     &  2.1     \\
\hline

 & $3/2^-$ &  0.277      &  0.00054$\pm0.21$ &  0.30 &  0.004 \\ 

 & $1/2^+$ &  (1.9)        &  $\simeq$0.7    &         &    \\ 

 & $5/2^-$ &  2.638$\pm5$   &  0.081$\pm$5  &  2.55   &  0.044  \\

 & $1/2^-$ &  3.11$^{\:c}$       &  3.1    &  2.73   &  1.0    \\

 & $5/2^+$ &  3.065$\pm30$     & 0.550$\pm$40 &  3.5    &  1.2  \\ 

 $^9$B & $3/2^+$ &     &      &  4.6    &  2.7  \\ 

 & $3/2^-_{2}$ &         &         &  4.2    &  1.4   \\

 & $7/2^-$  &   7.25$\pm60$    &  2.0$\pm$200  &  7.0    &  1.7    \\ 

 & $9/2^+$  &            &       &  6.6    &  3.3    \\

 & $5/2^-_{2}$ &         &       &  8.4    &  2.4    \\
\hline \hline

\end{tabular}

a) Ref.\cite{ajzen}. $\:\:\:$  b) Ref.\cite{dixit}. $\:\:\:$  c) Ref.
\cite{pugh}.

\vskip 1.5cm

\par\indent
The electromagnetic moments and the rms radii of proton, neutron, and 
nucleon, assuming pointlike nucleons, are included in Table 3. 
Bare operators are used in the calculation. 
The theory is found to reproduce the data very well. 
The fact that the total reaction cross section is reproduced well also 
supports  that the calculated ground state density is reliable. 
The $1/2^+\to 3/2^-$ $E1$ transition and the 
$5/2^-\to 3/2^-$ $E2/M1$ transitions are studied by 
treating the excited states as quasibound states. The calculated 
transition rates, shown in Table 3., are in good agreement with the 
experiments. 

\newpage

\begin{table}
\caption{Radii and electromagnetic properties of $^9$Be. The reduced 
matrix elements are given in Weisskopf units. The bare-nucleon 
charges and $g$-factors are used in the present calculation.  The effective 
charges were used in the shell model calculation of Refs. 
\protect{\cite{glickman}} and \protect{\cite{hees84}} to calculate 
the quadrupole moment and the $E2$ strength. See text for the 
$B(E1)$ value of the present calculation.}
\end{table}

\vspace {0.5mm}

\begin{tabular}{p{1cm}p{3cm}p{3cm}p{1.6cm}}
\hline\hline

 $J^{\pi}$ & & exp.$^a$ & present \\ \hline \hline

$3/2^-$ & $E$ (MeV)   &  $-$1.5735 & $-$1.431 \\
  & $r_m$ (fm)   &          &   2.50          \\
  & $r_p$ (fm)   &  2.37$\pm0.01$ &   2.39    \\
  & $r_n$ (fm)   &          &   2.58          \\
  & $\mu$ ($\mu_N$) & $-$1.1778$\pm$0.0009  & $-$1.169 \\
  & $Q$ (e fm$^2$) & 5.3$\pm$0.3 & 5.13                \\ 
  & $\sigma_R$ (mb)  & 825$\pm$20$^{\:b}$ &  850       \\ 
\hline

 $5/2^-$ & $E$ (MeV)  &  0.8559 & 0.883 \\
  & $B(E2;\frac{5}{2}^-$$\rightarrow$$\frac{3}{2}^{-})$  
  & 24.4$\pm$1.8 & 22.0 \\
  & $B(M1;\frac{5}{2}^-$$\rightarrow$$\frac{3}{2}^{-})$  
  & 0.30$\pm$0.03 & 0.229    \\ 
\hline 

 $1/2^+$ & $E$ (MeV)  & 0.111  \\
  & $B(E1;\frac{1}{2}^+$$\rightarrow$$\frac{3}{2}^{-})$ 
  & 0.22$\pm$0.09 & 0.24  \\
  &   &   &  0.18  \\
\hline 
\hline

\end{tabular}

\par\indent
The fact that the present calculation reproduced all the data very well 
strongly supports that the three-cluster model is quite appropriate for 
describing the structure of $^9$Be and $^9$B, provided that the three-body 
dynamics is treated properly in the calculation. 
Because the ground state and the $\frac{5}{2}^{-}$, 2.43 MeV state are 
described well by the present model, the $\beta$ decay of the $^9$Li ground 
state to these states is expected to further test the accuracy of their 
wave functions or an available wave function of $^9$Li. The experimental 
value of log$ft$ for the $\beta$ decay to the $^9$Be ground state is about 
5.31 \cite{ajzen,nyman}, indicating that the $\beta$-decay matrix element is 
fairly suppressed despite the allowed transition. The weak $\beta$ decay 
is ascribed to the fact that the spatial symmetry of the main component 
of $^9$Be is different from that of $^9$Li \cite{cohen}.
In fact the Gamow-Teller (GT) matrix element 
to any state of $^9$Be, if it is described by the $\alpha+\alpha+n$ 
three-cluster model, always vanishes regardless 
of the wave function of $^9$Li.  To explain the weak $\beta$ decay 
we have to admix a small component which is not taken into account in 
the three-cluster model. This possibility has been investigated by 
including the distortion of the $\alpha$-particle into $t+p$ and 
$h+n$ configurations. 
The calculated log$ft$ value turns out to be 5.60, indicating that 
we are on the right track. By being 
able to accommodate such distortion into the model consistently we have 
exemplified a unique advantage of the microscopic multicluster model.
\par\indent
We have investigated the question of whether or not some high isospin excited 
states of stable nuclei have extended halo-like structure.
For this we have focused on the 0$^+$ state of $^6$Li at 3.563 MeV excitation 
energy, which is the $T_z$=0 member of the isospin triplet together 
with the ground states of $^6$He and $^6$Be.
This indicates that the spatial structure of $^6$Li(0$^+$) is very similar 
to the $^6$He ground state which is known to have two-neutron halo structure.

We have done microscopic calculations for $^6$Li(0$^+$) with 
$\alpha$+$p$+$n$ three-body model by allowing the isospin mixing due to 
the Coulomb potential.
A large number of channels was taken into consideration in the calculation.
The isospin mixing was found to be moderate and does not destroy 
the isobaric analogue concept proposed in ref. \cite{suzu2} 
even near the three-particle threshold.
The accuracy of the wave function was tested by calculating the electron 
scattering form factors.
The calculated form factors are in fair agreement with experiment.

The neutron and proton density distributions of $^6$Li(0$^+$) are similar 
and more extended up to larger distance than the $^6$He density.
The matter size of $^6$Li(0$^+$) is calculated to be 2.73 fm, 
which is by 0.2 fm larger than the $^6$He size.
Our analysis strongly supports that the 3.563 MeV 0$^+$ state of $^6$Li 
has spatially extended halo-like structure formed by the neutron and 
proton outside the $\alpha$-particle. 
The inelastic proton \cite{proton} and pion \cite{pion} scatterings to 
this state show strong anomaly in the sense that any theoretical 
analysis leads to outstanding discrepancy from experiment. 
We think the consideration of the halo-like structure in this state 
is needed in such an analysis.

Further development of radioactive beam facilities will open up 
a spectroscopic study on high isospin excited states of stable nuclei 
as well as unstable nuclei near the drip-line and give us valuable information 
on the isospin impurity and the generalization of the neutron-halo concept.

\section{Comparision to other models}
As we have shown in a number of examples, the microscopic multicluster 
model provides us with a good and consistent description of light nuclei 
including unstable nuclei. The reason for success is that it can 
duly take into account the dynamical correlation between the nucleons 
as well as the asymptotic behavior characteristic to the weakly bound 
system. To treat the relative motion of the clusters flexibly enough, the 
trial function for the relative motion is chosen to be correlated Gaussians . 
The complexity of the wave functions 
is increased further if different cluster partitions are included in the 
model space. 
\par\indent
The direct comparison of our wave function and model space to that of
shell model or even to other versions of cluster models (which use
different trial functions) is very complicated. Some way for comparision,
hovewer, would be inevitably necessary to have deeper understanding of
applicability and limitations of different models. It would be useful  
to have  a simple measure to compare various types of wave functions. 
We have recently proposed to use the occupation probability of the number 
of total oscillator quanta as a possible candidate as a tool of 
comparison\cite{hoqt}. The probability $P_Q$ 
of a definite number of total HO quanta $Q$ can be obtained by 
calculating the expectation value of the operator $\cal O$
\begin{equation}
{\cal O}=\frac{1}{2\pi}\int_0^{2\pi}d\theta \ {\rm exp}\Bigl(
i\theta\ (\ \sum_{i=1}^A P_i\  
[H_{\rm HO}(i)-\frac{3}{2}]-Q)\Bigr), 
\end{equation}
where $H_{\rm HO}(i)$ is the 3-dimensional 
HO Hamiltonian divided by $\hbar\omega=\frac{2\hbar^2}{m}\gamma$. 

\newpage
\begin{table}
\caption{The occupation probability of the number of harmonic-oscillator 
quanta for microscopic 
multicluster-model wave functions. The probabilities for nucleons, 
protons, and neutrons are given in \% in the upper, middle, and lower 
rows, respectively, as a function of oscillator excitations. 
Asterisk indicates the probability of less than 1 \% and dashed line 
represents vanishing probability. The average number of 
oscillator excitations is given in the column labeled $<Q_{\rm exc}>$. 
The details of the wave functions are referred to Ref. [12] for $^6$He 
, to Ref. [13] for $^7$Li, $^8$Li, $^9$Li, and $^9$C, and to 
Ref. [14] for $^9$Be. }

\vspace {0.5mm}

\begin{tabular}{ccccccccccccccc}

state & {\rm rms radius}  &  \multicolumn{12}{c}{$Q_{\rm exc}$} & \\
\cline{3-14}

(model) & [fm] &
 0 & 1 & 2 & 3 & 4 & 5 & 6 & 7 & 8 & 9 &10 & 11 &
\raisebox{1.5ex}[0pt]{$<Q_{\rm exc}>$}  \\
\hline

$^6$He(0$^+$) & 
  $r_m=2.51$ & 60 & --- & 14 & --- & 12 & --- & 5  & --- & 3 
& --- & 2 & ---& 2.2 \\
($\alpha$+$n$+$n$) & $r_p\;=1.87$ & 74 & 10  & 11 & 2  & 1  & $\ast$ & $\ast$ 
& $\ast$ & $\ast$ & $\ast$ & $\ast$ & $\ast$ &
 0.5  \\
& $r_n\;=2.78$ & 67 & 3  & 8 & 5  & 7  & 2 & 2 & 1 & 1 & 
$\ast$ & $\ast$ & $\ast$ & 1.7   \\
\hline

$^7$Li(3/2$^-$) & 
  $r_m=2.34$ & 63 & --- & 20 & --- & 9  & --- & 4 & --- & 2 & --- &
$\ast$ & --- & 1.4  \\
($\alpha$+$t$) & $r_p\;=2.28$ & 77 & 2  & 16 & $\ast$ & 4 
& $\ast$ & $\ast$ & $\ast$ & $\ast$ & $\ast$ & $\ast$ &
$\ast$ & 0.6  \\
& $r_n\;=2.38$ & 73 & 1  & 17 & $\ast$ & 5  & $\ast$ & 1 & $\ast$ 
& $\ast$ & $\ast$ & $\ast$ & $\ast$ & 0.8  \\
\hline

$^7$Be(3/2$^-$) & 
  $r_m=2.36$ & 62 & --- & 20 & --- & 9  & --- & 4 & --- & 2 & --- &
 1 & --- & 1.6  \\
($\alpha$+$h$) & $r_p\;=2.41$ & 71 & 1  & 17 & $\ast$ & 5 
& $\ast$ & 2 & $\ast$ & $\ast$ & $\ast$ & $\ast$ &
$\ast$ & 0.9  \\
& $r_n\;=2.31$ & 75 & 2  & 16 & $\ast$ & 4  & $\ast$ & 1 & $\ast$ 
& $\ast$ & $\ast$ & $\ast$ & $\ast$ & 0.7  \\
\hline

$^8$Li(2$^+$) & 
  $r_m=2.45$ & 61 & --- & 18 & --- & 11  & --- & 4 & --- & 2 & --- & 1 & 
--- & 1.7  \\
($\alpha$+$t$+$n$)& $r_p\;=2.19$ & 79 & 6  & 11 & 1  & 2  & 
$\ast$ & $\ast$ & $\ast$  & $\ast$ & $\ast$ & $\ast$ & $\ast$ & 0.4  \\
& $r_n\;=2.60$ & 67 & 3  & 14 & 2  & 7   & 1 & 2 & $\ast$ & 1 & $\ast$
& $\ast$ & $\ast$ & 1.3  \\
\hline

$^8$B(2$^+$) & 
  $r_m=2.63$ & 54 & --- & 17 & --- & 12  & --- & 6 & --- & 3 & --- & 1 & --- &
2.7  \\
($\alpha$+$h$+$p$)& $r_p\;=2.83$ & 60 & 3  & 14 & 3  & 8  & 
 2 & 3 & 1  & 2 & $\ast$ & $\ast$ & $\ast$ & 2.1  \\
& $r_n\;=2.26$ & 74 & 9  & 12 & 2  & 2 & $\ast$ & $\ast$ & $\ast$ & $\ast$ & $\ast$
& $\ast$ & $\ast$ & 0.6  \\
\hline

$^9$Li(3/2$^-$) & 
  $r_m=2.40$ & 66 & --- & 17 & --- & 11  & --- & 4 & --- & 2 & --- & $\ast$
& --- & 1.3  \\
($\alpha$+$t$+$n$+$n$) & $r_p\;=2.10$ & 82 & 6  & 9 & 1  & 1 & $\ast$ 
& $\ast$ & $\ast$ & $\ast$ & $\ast$ & $\ast$ & $\ast$ & 0.4  \\
& $r_n\;=2.54$ & 71 & 3  & 12 & 2  & 6   & 1 & 2 & $\ast$ & $\ast$ 
& $\ast$ & $\ast$ & $\ast$ & 1.0  \\
\hline

$^9$C(3/2$^-$) & 
  $r_m=2.52$ & 60 & --- & 17 & --- & 12  & --- & 5 & --- & 3 & --- & 1
& --- & 1.8  \\
($\alpha$+$h$+$p$+$p$) & $r_p\;=2.68$ & 65 & 4  & 12 & 3  & 7 & 1 
& 2 & $\ast$ & 1 & $\ast$ & $\ast$ & $\ast$ & 1.4  \\
& $r_n\;=2.16$ & 79 & 8  & 9 & 2  & 1 & $\ast$ & $\ast$ & $\ast$ & $\ast$ 
& $\ast$ & $\ast$ & $\ast$ & 0.4  \\
\hline

$^9$Be(3/2$^-$) & 
  $r_m=2.50$ & 54 & --- & 21 & --- & 12  & --- & 5 & --- & 3 & --- &
2 & --- & 2.1  \\
($\alpha$+$\alpha$+$n$) & $r_p\;=2.39$ & 71 & 3  & 17 & 1  & 5 & $\ast$ & 1 
& $\ast$ & $\ast$ & $\ast$ & $\ast$ & $\ast$ & 0.8  \\
& $r_n\;=2.58$ & 65 & 2  & 18 & 1  & 8 & $\ast$ & 3 & $\ast$ & 1 
& $\ast$ & $\ast$ & $\ast$ & 1.3  \\

\end{tabular}
\end{table}

$P_i$ projects out either proton or neutron. It is set the unit 
operator when one calculates the number of total quanta occupied by both 
protons and neutrons. The advantege of this formalism is that 
the evaluation of the matrix element of this operator is very simple in cluster
models and it is trivial in shell model. Table 4. shows the 
examples for a pair of the mirror nuclei $^7$Li-$^7$Be, $^8$Li-$^8$B, 
and $^9$Li-$^9$C. The result for $^6$He and $^9$Be is also included 
in the table. The value of $\gamma$ is set 0.17 fm$^{-2}$ ($\hbar\omega$
=14.4 MeV). The probabilities 
are given as a function of $Q_{\rm exc}=Q-Q_{\rm min}$, where 
$Q_{\rm min}$ is the minimum number of HO quanta for the 
lowest Pauli-allowed configuration. The lowest 0$\hbar
\omega$ component is around 50-60 \% for most cases and the sum of 
0, 2, and 4$\hbar\omega$ components accumulates to about 90 \%. The 
admixtures of higher components than $Q_{\rm exc}=4$ are significant in 
the ground states of $^8$B and $^9$Be and also in the ground 
state of $^6$He.  

\section{Extension to larger systems}

The application of the SVM on correlated Gaussian basis for more 
than $N$=6-7 body system is difficult as both the partial wave expansion 
and the calculation of the matrix elements 
of the Hamiltonian become too complicated and computer time consuming. 
Let us try to use some other type of basis functions to avoid these problems.
The Gaussian packet $\varphi^{\gamma}_{\bf s} ({\bf r})=(2\nu/\pi)^{3/4} 
{\rm exp} \lbrace -\gamma({\bf r}-{\bf s})^2 \rbrace$ functions 
are also often used in few-body and few-cluster calculations. These functions
do not belong to a particular orbital angular momentum and if one uses them 
as single particle basis functions in a Slater determinant, the calculation 
of the matrix elements will not cause serious problems. 
In a variational calculation the basis functions take the form:
\begin{equation}
\Phi_{\bf s}={1\over \sqrt{N!}}{\rm det}\lbrace 
\varphi_{{\bf s}_i}^{\nu_i}({\bf r}_j) \rbrace
\end{equation}
The wave function of the system is approximated by linear combinations
of these Slater determinants.
States of good angular momenta can be obtained by letting an angular momentum 
projection operator act on the full wave function. 

\begin{table}[t]
\caption{The binding energies (in MeV) of different A-nucleon 
systems interacting via the Volkov potential ($m=0.6$)} 
\begin{tabular}{|c|c|c|c|c|}
\hline 
        &   $K=1$    &   $K=100$   &   method   &    result  \\
\hline 
$^3$He  &  $ -6.66$  &   $ -8.31$  &   SVM      &   $-8.46$  \\
$^4$He  &  $-27.92$  &   $-29.75$  &   SVM      &   $-30.42$ \\
$^6$He  &  $-23.59$  &   $-29.38$  &   SVM      &   $-31.82$ \\
$^8$Be  &  $-52.60$  &   $-57.09$  &            &            \\
$^{16}$O&  $-1100.1$ &   $---$     &   IDEA     &   $-1101.$ \\
\hline   
\end{tabular}
\end{table}
The adequate choice of the nonlinear parameters 
$\lbrace \nu_i,{\bf s}_i\rbrace_{i=1}^N$ of the Slater determinants
is very important. These nonlinear parameters can be selected 
by the stochastic variational method (SVM) or by an appropriate  
direct optimatization. To keep the cost of the optimatization low only a set
of the nonlinear parameters in a given Slater determinant was optimized at a 
time, while those of the others were kept fixed.
\par\indent
The energy quite slowly 
converge on this uncorrelated basis and for few-body systems it fails 
to reach the same value as on the angular momentum projected Gaussian 
(see Table 5.). For heavier systems, where the angular momentum projected 
Gaussian basis is not feasible, 
the results on shifted Gaussian basis is quite close to those of other 
methods such as the Integrodifferential Equation Approach (IDEA) or the 
Hartree-Fock-Bogoljubov (HFB) method. 
\par\indent
The Antisymmetrized Molecular Dynamics (AMD) \cite{AMD} and Fermionic 
Molecular Dynamics (FMD) \cite{FMD} methods use the shifted Gaussian 
basis functions and often approximate
the wave function of a multinucleon system by only one Slater-determinant 
($K=1$ in Table 5.). Table 5. shows that this approximation can be 
insufficient, and the linear combination  of Slater determinants 
considerably lowers the ground state energy.  

\section{Summary}

In summary, we presented the new results of the microscopic multicluster 
model using the stochastic variational method. The stochastic variational 
method provides us with accurate energies and wave functions. We have 
determined physical quantities of interest (proton, neutron and matter
distributions and radii, magnetic and quadrupole moments, electromagnetic
transition rates, beta-decay probabilities, momentum distribution of
fragments, interaction cross sections, spectroscopic amplitudes) for
various light nuclei. We have discussed the relation of our approach 
to other models. Further applications for light nuclei, for example
$^{10}$Be, $^{11}$Be, $^{11}$Li are under way.
\vskip 1.cm
This work was supported  by 
Grant-in Aids for Scientific Research on Priority Areas (No. 05243102), 
for Scientific Research (C) (No. 0664038), and for International 
Scientific Research (Joint Research) (No. 08044065) of the
Ministry of Education, Science and Culture (Japan) and by OTKA Grant No. 
T17298 (Hungary). Most of the calculations were done with the use of 
RIKEN's VPP500 computer.

\end{document}